# Deep Learning Enabled Real-Time Photoacoustic Tomography System via Single Data Acquisition Channel

Hengrong Lan[1,2,3,†], Daohuai Jiang[1,2,3,†], Feng Gao[1], and Fei Gao[1,*]

[1] *Hybrid Imaging System Laboratory, Shanghai Engineering Research Center of Intelligent Vision and Imaging, School of Information Science and Technology, ShanghaiTech University, Shanghai 201210, China*
[2] *Chinese Academy of Sciences, Shanghai Institute of Microsystem and Information Technology, Shanghai 200050, China*
[3] *University of Chinese Academy of Sciences, Beijing 100049, China*
[†] *equal contribution*
[*] *gaofei@shanghaitech.edu.cn*

*Abstract*—Photoacoustic computed tomography (PACT) combines the optical contrast of optical imaging and the penetrability of sonography. In this work, we develop a novel PACT system to provide real-time imaging, which is achieved by a 120-elements ultrasound array only using a single data acquisition (DAQ) channel. To reduce the channel number of DAQ, we superimpose 30 nearby channels' signals together in the analog domain, and shrinking to 4 channels of data (120/30=4). Furthermore, a four-to-one delay-line module is designed to combine these four channels' data into one channel before entering the single-channel DAQ, followed by decoupling the signals after data acquisition. To reconstruct the image from four superimposed 30-channels' PA signals, we train a dedicated deep learning model to reconstruct the final PA image. In this paper, we present the preliminary results of phantom and in-vivo experiments, which manifests its robust real-time imaging performance. The significance of this novel PACT system is that it dramatically reduces the cost of multi-channel DAQ module (from 120 channels to 1 channel), paving the way to a portable, low-cost and real-time PACT system.

## I. INTRODUCTION

As a new imaging modality, photoacoustic computed tomography (PACT) has emerged to show great potential in biomedical imaging areas, which is based on photoacoustic (PA) effect generating ultrasound by a nanosecond pulsed laser [1-4]. It blends the spatial resolution of ultrasound imaging and the high contrast of spectroscopic optical absorption. The ultrasonic detectors are placed around the object to receive the PA signals simultaneously. Then, a reconstruction algorithm is used to recover the initial pressure distribution [5-7]. In recent years, PACT has been applied in many preclinical or clinical applications, such as small animal functional imaging and early breast tumor detection [8-19].

Several PACT systems have been developed to image the specific tissues (e.g. breast) or the whole body of small animals in real-time, which requires high performance toward preclinical or clinical application [20-25]. Specifically, the reported PACT systems are mainly improved in the following ways: (1) Increasing the number of transducers in planar or hemispherical array scheme. Gamelin et al. proposed a real-time PACT system with 512-elements for small animals' imaging [21]; Lin et al. developed a single-breath-hold PACT breast imaging system delivering high image quality with 512-elements ultrasound probe [22]; (2) Combining other imaging modality (e.g. ultrasound imaging) with PACT in linear array scheme. Park et al. presented a PA, ultrasound, magnetic resonance triple-mode system [23]; (3) Reducing the total system cost. SK Kalva et al. introduced a fast imaging PACT system based on pulsed laser diode (LD) [24]; Zafar et al. developed a low-cost scanning system using just a few detectors [25]. However, reducing the number of detectors and DAQ channels may result in more deficient image quality damage or slower imaging speed (mechanical scanning is required). Recently, deep learning has been dramatically developed in signals, images, and video processing. Deep learning methods have cut a figure in PA image reconstruction problems from raw data or imperfect image [26-32].

Some works have presented the single detector PACT system by other physical principles, which significantly reduced the number of signals required for reconstructing PA image. The single-channel approach has made significant progress in the PA imaging field in recent decades, promoting real-time imaging potential [33-36]. For instance, Y. Li et al. achieved a wide-field imaging system through an ergodic relay [35]. Furthermore, compressed sensing boosts the progress of single detector system. In [33], a single-pixel camera PAT is demonstrated, which implements a 3D imaging system with compressed sensing. In [34], the authors introduced a single detector PACT system using acoustic scattered encoding. Furthermore, Yuning Guo et al. also proposed a single-channel PACT with compressed sensing [36]. This paper attempts to report a novel low-cost PACT system from a different perspective, which achieves real-time imaging performance via only single-channel DAQ. The DAQ is placed after PA signals' superimposition and 4-to-1 delay-line module in the analog domain, followed by deep learning based image reconstruction from the superimposed and delayed PA signals. Our proposed system combines minor hardware with a novel DL framework, which does not change imaging setup compared with previous single-channel systems.

## II. METHOD

### A. Overview

The single-channel PACT system is shown in Fig. 1: A computer controls a pulsed laser, and a 120-channels ring-shaped transducer array can real-timely receive the 120-channels' PA signals fed into an analog adder module, is placed in the water tank.

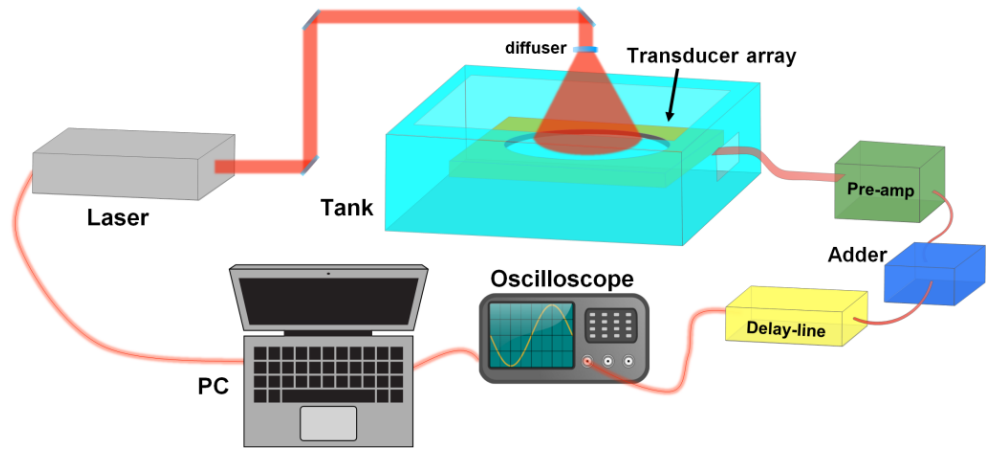

Fig. 1. The overview of the proposed single-channel PACT system. PC: personal computer. Pre-amp: Pre-amplifier

The PA signals are superimposed every 30 channels' PA signals into one composite signal after pre-amplification, and then 120-channels' PA signals are reduced to four channels. For the adder module, we use an analog addition circuit with 30 inputs and one output, which matches the bandwidth with the US transducer. The four superimposed PA signals go through a 4-to-1 delay-line module, which can adequately delay the four composite PA signals and sum them into one combined signal (Preliminary results about the 4-to-1 delay-line module can be found in [37, 38]). We summarize the flow diagram of the system operation in Fig. 2. 120-channels PA signals can be detected by transducers, feeding them into the adder module that superimposes them into four composite signals. The mixed PA signals are fed into the delay-line module, combining four channels into one combined PA signal with different time delay. By playing the 4-to-1 delay-line module, the four parallel signals are covert into one mixed serial signals. Then a single-channel DAQ converts this combined PA signal to digital data. In the digital domain, the combined signals that with different time delay can be recovered into four without time delay composite PA signals via signal processing, which can be used to reconstruct the final PA image by dedicated deep learning framework. The delay-line module and deep-learning based reconstruction framework will be introduced in the following sessions.

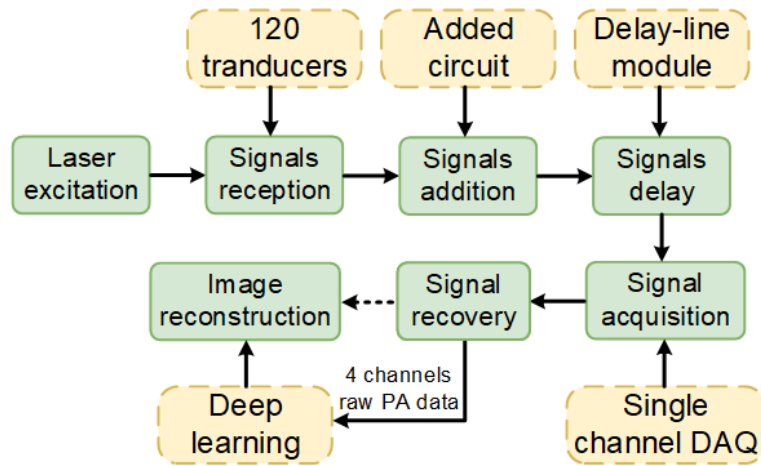

Fig. 2. The flow diagram of the operation for proposed single-channel PACT system.

## B. Four-to-one delay-line module

The four-to-one delay-line module is to merge 4 PA signals into one channel properly. To achieve this, we proposed the time-sharing multiplex transmission method for the PACT system. The PA signal is intrinsically a very short ultrasound pulse, whose duration is usually less than 50 microseconds for a ring setup PAT system. A scanning radius less than 75 millimetres (assuming that acoustic wave transmits speed is 1.5 mm/$\mu$s). Therefore, the four PA signals with different and sufficient time delays (e.g. 0 $\mu$s, 50 $\mu$s, 100 $\mu$s, 150 $\mu$s) can be distinguished and merged into one combined signal, which can be recovered in the digital domain. To achieve tens of microseconds time delay for analog pulse signals, we proposed and fabricated a four-to-one delay-line module based on acoustic delay method.

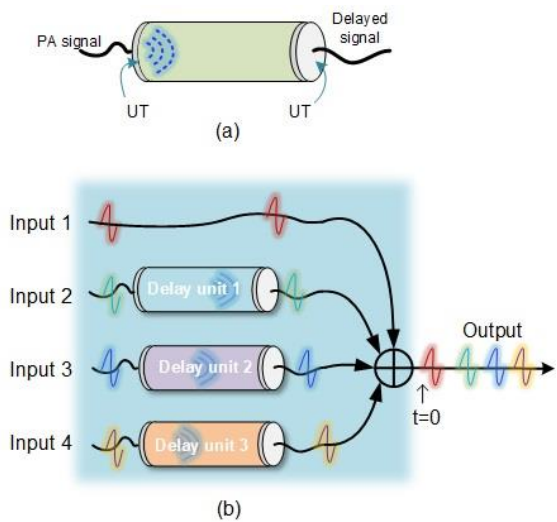

Fig. 3. The schematic of the four-to-one delay-line module, (a) the structure of the delay unit, UT: ultrasonic transmitter, UR: ultrasonic receiver, TM: ultrasound transmission medium, (b) the structure of the four-to-one delay-line module.

The schematic of the four-to-one delay-line module is shown in Fig. 3, including three delay units and a multi-channel adder in Fig. 3(b). Fig. 3 (a) shows the delay unit structure, which includes two ultrasound transducers to transfer the signals between electrical and ultrasound modes, i.e. one for transmitting and another for receiving. A low noise amplifier is used for amplifying the composite PA signal before entering the delay line unit. After signal amplification, the PA signal is converted into an acoustic signal by an ultrasound transducer, propagating in the transmission medium. At the end of the transmission, the acoustic signal is again converted back into an electrical signal by another transducer. This research selected water as the ultrasound transmission medium considering its quite low acoustic attenuation over a wide bandwidth and relatively small velocity (~1500 m/s). The central frequency and bandwidth of the ultrasound transducer in the delay line need to be matched to PA signals that can minimize distortion of the delayed signal.

The proposed four-in-one acoustic delay-line module is based on the abovementioned delay line unit. The structure of the module shows that there are four inputs and one output. The input signals are transmitted by the module with different time delays and combined into one output by an analog signal adder. The parallel input signals are almost synchronous to enter the delay-line module. To discriminate different input signals after summation, these input signals should be transmitted with varying delays of time. Specifically, the first input signal is connected to the analog adder directly without delay; the second to fourth input signals are delayed differently. To avoid aliasing of the delayed echo signals in the time domain, each delay unit's delay time can follow the relationship as shown in Table 1. Where $T$ stands for the delay period, b stands for delay time bias, and it is non-negative. The first echo signal of unit 2 is still larger than the input 4's delayed signal. Therefore, all the echo signals do not alias with the delayed four input signals. Finally, the three delayed signals from input 2~4 and a non-delayed signal from input one are combined into one output by the analog adder.

**Table 1** Delay time and first echo time

| Input | Delay time | 1st echo time |
|-------|------------|---------------|
| 1     | 0          | none          |
| 2     | $1.5T+b$   | $4.5T+3b$     |
| 3     | $2.5T+b$   | $7.5T+3b$     |
| 4     | $3.5T+b$   | $10.5T+3b$    |

The delay time depends on the length of the ultrasound transmission medium. The relationship between the length of the transmission medium and delay time can be calculated easily with the fixed ultrasound propagation speed in this medium. Fig. 3 (b) shows the structure of the four-to-one delay-line module, the delay time of each delay unit is different and constant. By applying this module, the four coinstantaneous PA signals with varying time delays can be merged into one output by an analog signal adder. In this work, the delay time of each unit is 50, 100 and 150 microseconds, respectively.

The composite signal can be reconstructed into four independent signals with the time-shifting operation in the digital domain. The signal reconstruction can be separated into two steps basically. As shown in Fig. 4, Fig. 4(a) is the four-to-one composite signal, (b) shows the separated signals and Fig. 4(c) are the reconstructed signals after the time-shifting operation.

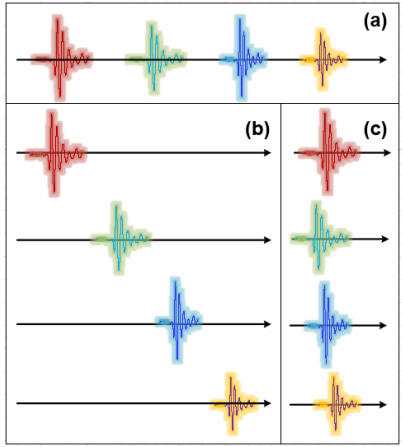

Fig. 4. The separation process for the delay-line module. (a) the one channel four-to-one PA signal; (b) four channels separated delayed PA signals; (c) four channels recovered PA signals.

## C. Deep learning reconstruction architecture

The deep learning architecture for PA image reconstruction is shown in Fig. 5, which takes four superimposed PA signals from the abovementioned delay-line unit as input and generates the reconstructed PA image. As shown in Fig. 5, an encoder comprises a long short-term memory (LSTM) and a full connection layer, which encode input signals to 64 feature sizes before the decoder. It is noteworthy that the 64 feature size needs reshape to $8 \times 8$ size before taking it into the decoder. Four up-sampled layers comprise an up-sampled operation for the decoder and two convolutions, batch normalizations, and leaky Rectified Linear Unit (ReLU) operations. Afterwards, we can obtain the final image through a Residual-block (Res-block) [39].

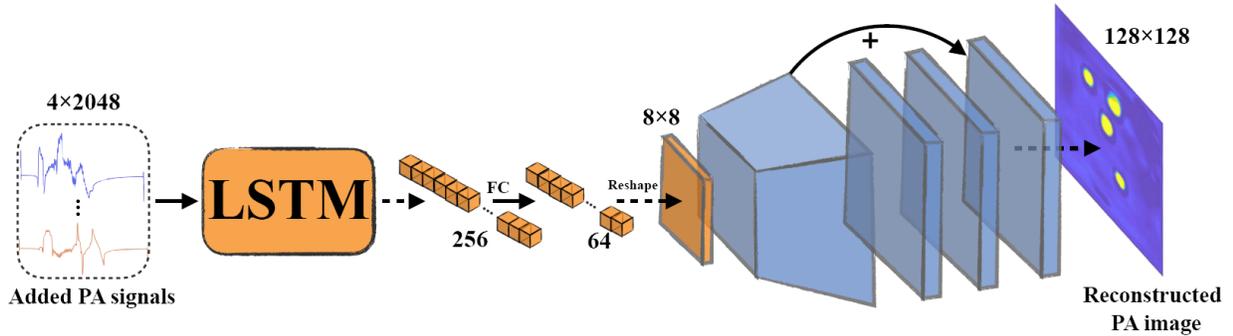

Fig. 5. The overview of proposed deep learning architecture, the superimposed signals are converted to semantic features by LSTM and reshaped as 256 features. The final three blue features contain a residual operation. LSTM: long short-term memory, FC: full connection.

Considering that the input size has an extreme asymmetry, including four size spatial channels and 2048 size temporal distribution, we apply a recurrent neural network (RNN) to process the spatial-temporal data and extract the semantic information to full connection layers encode the semantic features.

The semantic feature from the encoder is fed into the decoder after a reshaping operation. The decoder converts the semantic features to an image, which is composed of four up-sampled layers as follows:

$$UP(x) = \text{ReLU}\{w_2 * \text{ReLU}[w_1 * up(x)]\}, \tag{1}$$

where $up(\cdot)$ is an up-sampled operation, $w_1$ and $w_2$ are the weight of two convolutions, and we use leaky ReLU as activation function that can be expressed as:

$$f(x) = \begin{cases} x, & x > 0 \\ \lambda \cdot x, & x \leq 0 \end{cases}, \tag{2}$$

where $\lambda$ is the coefficient of leakage, equal to 0.2 in this work. After that, the image features are fed into a Res-block, which is expressed as follow:

$$\text{Res}(x) = \text{ReLU}\{w_3 * [x + w_2 * \text{ReLU}(w_1 * x)]\}. \tag{3}$$

Considering that deep learning is a data-driven method, we need plenty of training data. But the current PAI equipment is still not available in the clinic, and we have to use synthetic data generated by MATLAB toolbox k-Wave [40]. We use MSE function to minimize the error in the training, which is expressed as follow:

$$L_{rec}(y) = \frac{1}{2}\|y - gt\|_F^2, \tag{4}$$

where $gt$ and $y$ denote ground-truth and output image, respectively.

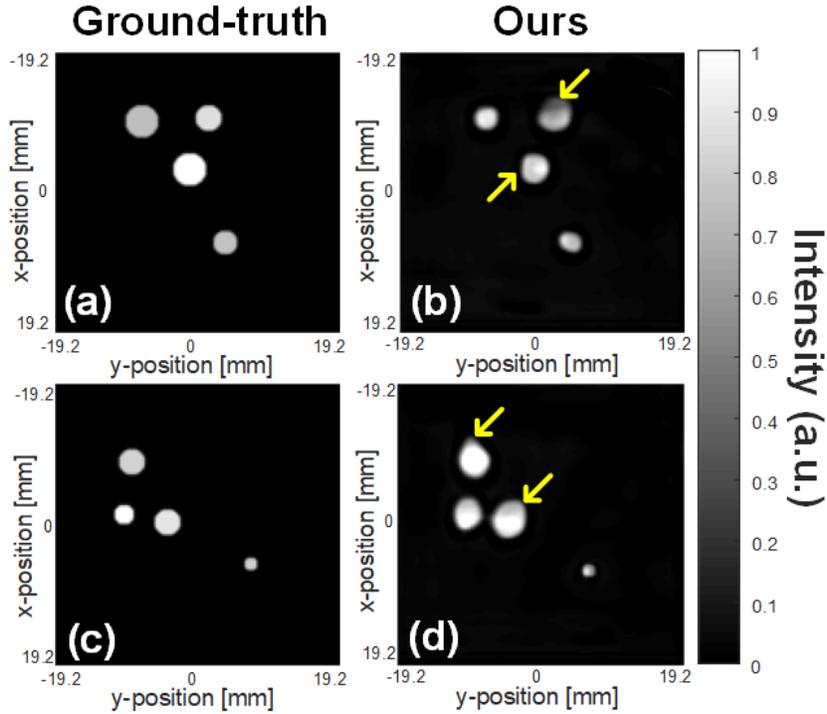

Fig.6. The reconstructed results of test samples. (a) ground-truth of sample 1; (b) reconstructed results of sample 1; (c) ground-truth of sample 2; (d) reconstructed results of sample 2. The yellow arrows indicate the blurs and deformations of the results.

In this paper, Pytorch is used to implement the deep learning method. The network is trained on the hardware platform, consisting of two Intel Xeon E5-2690 (2.6GHz) CPUs and four NVIDIA GTX 1080Ti graphics cards. The batch size is set as 64, and the initial learning rate is 0.005. The optimization algorithm we select in this paper is Adam [41].

## III. EXPERIMENTS

The deep-learning-based method requires plenty of data for training, so we first validate the reconstruction method with the simulation in session III.A. And then, we need to train our deep learning model with a set of *in-vivo* fish data and then demonstrate our system on the *in-vivo* experiment.

### A. Simulated experiment for deep learning model

We use MATLAB toolbox k-Wave to generate the data, which uses numerical phantoms consisting of four discs. The discs are randomly placed in the region of interest (ROI) within 38.4×38.4 mm area, and the size of the disc is randomly set from 0.75mm to 2.25mm. 120 sensors are evenly placed as a circle, whose center frequency is 7.5 MHz with 80% fractional bandwidth. The speed of ultrasound is 1500 m/s in soft tissue. The transducers can receive 2048 points data for every channel, and then we superimpose 30 channels' data, leading to four channels' superimposed PA signals. Namely, every data can be allotted 4×2048 size. Finally, we obtain 4500 training data and 1000 test data.

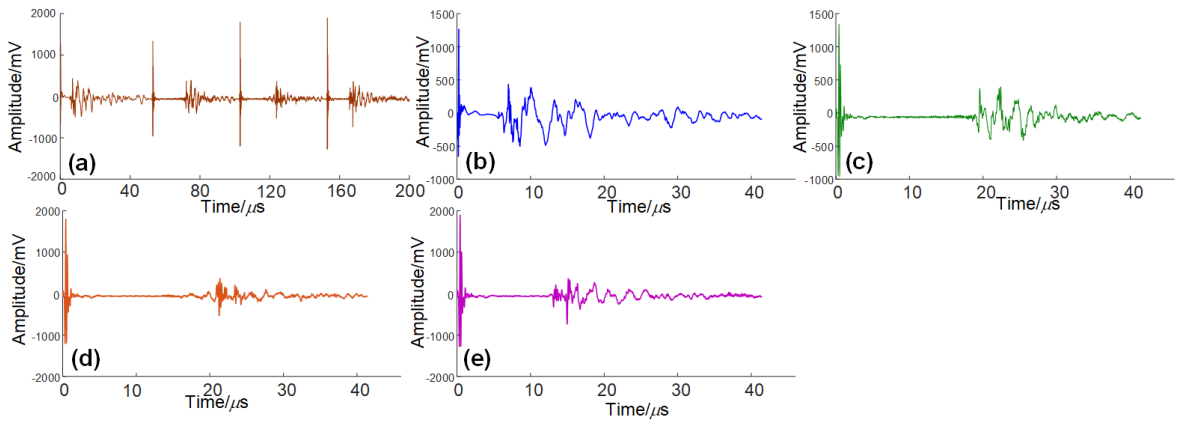

Fig. 7. (a) PA data received by one-channel DAQ; (b) recovered first superimposed PA data; (c) recovered second superimposed PA data; (d) recovered third superimposed PA data; (e) recovered fourth superimposed PA data.

### B. In-vivo fish experiment

We use the *in-vivo* fish PA image data to train the deep learning model. A set of training data should be acquired by a general PACT system and used to train the model. 120 elements transducer (central frequency: 7.5 MHz, Doppler Inc.) is placed surrounding the target with a 30 mm radius. The pitch size is about 1.5 mm. The pulsed laser (532 nm wavelength, 10 Hz repetition rate) is used to illuminate the sample covering ROI. The power of output conforms to the limitation of ANSI (20 mJ/cm$^2$ for 532 nm wavelength). A data acquisition card (LEGION ADC, PhotoSound) received and amplified the PA signals with 40 MSa/s sampling rate. Finally, we obtained 1900 samples, and we picked 1500 for training, 400 remaining for testing.

We further demonstrate the *in-vivo* fish imaging experiment using our proposed PACT system, as shown in Fig. 1. The transducer is the same with the general PACT system. Only one acquisition channel of an oscilloscope (DPO5204B, Tektronix) is used to collect the PA data after amplification (AMP16t, PhotoSound), superimposition operation, and delay-line module.

## IV. RESULTS

### A. Simulation results

We show two samples' results of the test set in Fig. 6, containing four discs for every sample. Fig. 6(a) and Fig. 6(c) are the ground-truth of these samples; Fig. 6(b) and Fig. 6(d) are reconstructed results by the proposed deep learning framework, respectively. The results show some blurs and deformations, as the yellow arrows indicated in Fig. 6(b) and 6(d). Still, the locations and size are maintained quite well with satisfactory contrast and resolution.

### B. In-vivo experimental results

To illustrate the delay-line module, we plot the one channel raw PA data and recovered four channels' PA data in Fig. 7. One channel superimposed and delayed PA data received by DAQ is shown in Fig. 7 (a), which has a sufficient 50 $\mu s$ delay time for every channel. Fig. 7(b)-(e) are recovered PA data from Fig. 7 (a) of the four channels, and every channel's PA data indicates the superimposition of 30 channels' raw PA signals.

The fish imaging results are shown in Fig. 8. Fig. 8(a)-(c) show three different results of the trained model, and Fig. 8(d)-(f) show the ground-truth of these results obtained by full 120 channels' data. It shows that the artifacts and low value in ground-truth cannot be reconstructed since the superimposed signal of multi-channel data decreases the weight from four channels data. It may be a limitation of our reconstruction method. However, it still can distinguish the outline of the fish from Fig. 8(a)-(c) with much fewer artifacts. We demonstrate the *in-vivo* experiment at 10 Hz frame rate (video provided in supplementary material), which is limited by the low repetition rate of the laser. The theoretical frame rate can be up to 30 Hz if using a high-repetition-rate pulsed laser.

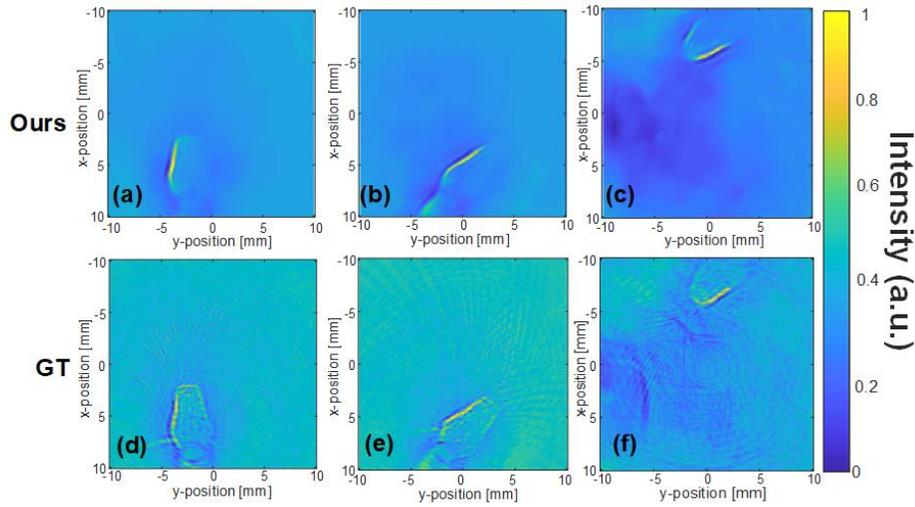

Fig.8. The PA imaging results of *in-vivo* fish. (a-c) The reconstructed results of three samples, (d-f) The ground-truth of three samples. GT: ground-truth.

We can further compare the time consumption of our proposed PACT system with a conventional single-channel PACT system (e.g. [17]). A non-focused ultrasound transducer rotates around the sample by a rotary motor to detect the PA signals. An amplifier is used to connect with the transducer to amplify the PA signals, and an oscilloscope acquires the PA data transferred to a computer. Our proposed system replaces the single detector with an array transducer and adds corresponding amplification channels compared with the conventional single-channel system. A superimposition module and delay-line module are added between the amplification and acquisition. We divide the operation procedure into data acquisition and image processing. Data acquisition includes the PA signal detection and processing before entering the DAQ; image processing includes signals recovery and image reconstruction after the PA signal is digitized by DAQ. For conventional single-channel PACT system, the transducer needs to mechanically rotate at 120 positions for PA signal detection repeatedly, which is quite time-consuming (261 seconds for 120 positions' PA signal detection in Table. 1). The image reconstruction of the conventional PACT system is by delay-and-sum (DAS) algorithm. On the other hand, our proposed PACT system acquires all the PA data within 2.35 ms using only single-channel DAQ, followed by deep-learning based image reconstruction algorithm that is much faster than DAS (28 ms v.s. 159 ms). Moreover, we also compare the time consumption of the conventional rotary scanning single-channel PACT system with our proposed system. By calculating the total time consumption shown in Table. II, our proposed PACT system is nearly 8600 times faster than the conventional single-channel PACT system. However, the practical time consumption could be limited due to other reasons, such as the repetition rate of the laser. It can be improved by using a light source with a high repetition rate.

It shows the great potential of our proposed PACT system for real-time PA imaging with significantly lower DAQ cost using a single channel. Last but not least, the quality of the reconstructed PA image by the deep-learning based algorithm shows much fewer artifacts compared with the conventional DAS algorithm (Fig. 8(c)).

TABLE II. The time consumption of the proposed system and the conventional single-channel PACT system.

| Time consumption | *Our proposed single-channel PACT system* | *Conventional single-channel PACT system* |
|---|---|---|
| Data Acquirement | 2.35ms | 261.6s |
| Image Processing | 28ms | 159ms |
| Total | 30.35ms | 261.759s |

## V. Discussion

In our system, the deep learning model is used to reconstruct the PA image from four channels' PA signals, which provides an effective and low-cost method for single-channel real-time imaging (single channel combined with the 4-to-1 delay-line module). It shows a high-speed reconstruction performance due to the powerful computation capability of GPU. On the other hand, it is impossible to reconstruct the image by the conventional method from the 4 superimposed signals. However, some issues of this approach deserve further improvement. For instance, our reconstruction method is not sensitive to the size of the disc-like target; artifacts and small values in ground-truth image could be ignored and hidden from the *in-vivo* experimental results. Namely, the artifacts can be removed in experiments, which could be beneficial to us sometimes. All these issues will be improved in our future work.

Our proposed system could be similar to the widely studied PACT systems except for the data acquisition procedure, which occupies the major cost of the imaging system except for the laser source. Furthermore, multi-channel DAQ does not have a cheap alternative, unlike the laser source (e.g. LED). Compared with widely studied real-time PA imaging system with hundreds of channels, the cost of our system could be closer to single-channel system. On the other hand, compared with single-channel PA imaging system, our proposed system speeds up the imaging rate achieving real-time performance. Therefore, our system provides a trade-off scheme between real-time imaging requirements and the expensive cost of multi-channel DAQ.

There is a trade-off between hardware cost (less channel number gives less hardware cost) and training cost (less channel number gives higher training cost). However, the training cost can be ameliorated by the fast-developing GPU computation power. Therefore, it could be a good trade-off by selecting 4 channels' data to train the network while maintaining a low hardware cost. Meanwhile, we could recover more channel data by improving the delay-line module in the future, providing more features to reconstruction with deep learning or general algorithms.

Deep learning-based reconstructions have different schemes, and our scheme is that map the function from raw sensor data to the final image directly. We reduce the dependence on time-delay in this work, and this scheme learns the distribution of target from data. For instance, T. Tong et al. proposed to cut the valid data that ensures the input data are close to square by sacrificing the time-delay information [42].

Our system has different advantages compared with other systems: 1. our system dramatically improves the speed imaging and achieves real-time imaging performance compared with conventional scanning-based single-channel detection; 2. compared with conventional multi-channel detection, our system dramatically decreases the cost from 120 channels to 1 channel; 3. compared with the new schemes recently proposed in [34, 36], we achieve higher image quality and faster imaging speed with more flexibility for real clinical applications.

## VI. CONCLUSIONS

In this paper, we developed a novel low-cost real-time PACT system that collects 120 channels' PA data by only a single data acquisition channel. 30-to-1 superimposition can decline the 120 channels' PA data to four channels, then four-to-one delay-line module can further combine the four channels into one channel PA data. To prove the feasibility of the system, we can reconstruct the disc-like phantom using four superimposed PA signals benefiting from our proposed deep learning approach. The phantom result has been demonstrated using our proposed PACT system and shows a robust performance with much fewer artifacts compared with the conventional PACT system. Furthermore, we can implement real-time imaging, which is never achieved in conventional single-channel PACT system. In the future work, we will further improve the system using an economical laser source for even lower cost and apply our system to the vessel or other *in-vivo* imaging applications.